Applied Physics Letters

ACCEPTED MANUSCRIPT

This is the author's peer reviewed, accepted manuscript. However, the online version of record will be different from this version once it has been copyedited and typeset.

PLEASE CITE THIS ARTICLE AS DOI: 10.1063/5.0062106

**Substrate impact on the thickness dependence of vibrational and optical properties of large area MoS$_2$ produced by gold-assisted exfoliation**

S. E. Panasci[1,2], E. Schilirò[1], F. Migliore[3], M. Cannas[3], F. M. Gelardi[3], F. Roccaforte[1], F. Giannazzo[1]* and S. Agnello[3,1,4].

[1] Institute for Microelectronic and Microsystems (CNR-IMM), Z.I. VIII Strada 5, 95121, Catania, Italy

[2] Department of Physics and Astronomy, University of Catania, Via Santa Sofia 64, 95123 Catania, Italy

[3] Department of Physics and Chemistry Emilio Segre', University of Palermo, Via Archirafi 36, 90123 Palermo, Italy

[4] ATeN Center, University of Palermo, Viale delle Scienze Ed. 18, 90128 Palermo, Italy

*E-mail: filippo.giannazzo@imm.cnr.it

## Abstract

The gold-assisted exfoliation is a very effective method to produce large-area (cm$^2$-scale) membranes of molybdenum disulfide (MoS$_2$) for electronics. However, the strong MoS$_2$/Au interaction, beneficial for the exfoliation process, has a strong impact on the vibrational and light emission properties of MoS$_2$. Here, we report an atomic force microscopy (AFM), micro-Raman (μ-R) and micro-Photoluminescence (μ-PL) investigation of 2H-MoS$_2$ with variable thickness exfoliated on Au and subsequently transferred on an Al$_2$O$_3$/Si substrate. The E$_{2g}$ - A$_{1g}$ vibrational modes separation Δω (typically used to estimate MoS$_2$ thickness) exhibits an anomalous large value (Δω≈21.2 cm$^{-1}$) for monolayer (1L) MoS$_2$ on Au as compared to the typical one (Δω≈18.5 cm$^{-1}$) measured on 1L MoS$_2$ on Al$_2$O$_3$. Such substrate-related differences, explained in terms of tensile strain and p-type doping arising from the MoS$_2$/Au interaction, were found to gradually decrease while increasing the number of MoS$_2$ layers. Furthermore, μ-PL spectra for 1L MoS$_2$ on Au exhibit a strong quenching and an overall red-shift of the main emission peak at 1.79 eV, compared to the 1.84 eV peak for 1L $MoS_2$ on $Al_2O_3$. After PL spectra deconvolution, such red shift was explained in terms of a higher trion/exciton intensity ratio, probably due to the higher polarizability of the metal substrate, as well as to the smaller equilibrium distance at $MoS_2$/Au interface.

In the last years, molybdenum disulfide ($MoS_2$) have been widely investigated, due to the broad range of potential applications in the fields of optoelectronics, nanoelectronics, sensing and energy [1,2,3,4]. Several synthesis methods of $MoS_2$ films have been explored so far, including top-down and bottom-up methods [5]. While the highest electronic quality $MoS_2$ is still produced by mechanical exfoliation, the micrometer size of the flakes obtained by this approach makes it unsuitable for practical applications. In this context, gold-assisted mechanical exfoliation has recently received increasing attention as an effective method to separate large area ($cm^2$-scale) $MoS_2$ with excellent electronic quality from molybdenite crystals [6,7,8,9]. Since the interaction between sulfur and Au atoms [10] is stronger than the interlayer Van der Waals (VdW) bonds in the layered crystal, ultra-thin membranes (predominantly composed by monolayer (1L) $MoS_2$, but also containing bilayer (2L) and few-layer (FL) regions) are obtained simply pressing a bulk $MoS_2$ stamp on a clean Au surface. These membranes can be subsequently transferred to insulating or semiconductor substrates to fabricate electronic/optoelectronic devices, showing performances comparable to those obtained with the best quality mechanically exfoliated $MoS_2$ [7]. Furthermore, as-exfoliated 1L $MoS_2$ on Au electrodes have been employed for memristor applications [11]. Finally, the Au-assisted exfoliation has been recently extended to a large number of layered crystals beyond $MoS_2$, including other transition metal dichalchogenides ($MoSe_2$, $MoTe_2$, 1T-$MoTe_2$, $WS_2$, $WSe_2$, $WTe_2$, $TiS_2$, $TiSe_2$, $IrTe_2$, $SnS_2$, $SnSe_2$, $NbSe_2$, $NbTe_2$, $VSe_2$, $TaS_2$, $TaSe_2$,

PdSe$_2$), metal monochalcogenides (e.g., GaS), black-phosphorus, black-arsenic, metal trichlorides (RuCl$_3$), and magnetic compounds (Fe$_3$GeTe$_2$) [12]. Hence, it represents a powerful method for the realization of artificial vdW heterostructures [13,14,15] and hybrid 2D/bulk semiconductor devices [16,17,18].

The strong MoS$_2$/Au interaction, which is beneficial for the large-area exfoliation process, has a strong impact on the electronic, vibrational and light emission properties of MoS$_2$. Different studies have been reported on the strain and doping of 1L MoS$_2$ induced by the gold substrate [6,9]. In this context, investigating the Au substrate effects on MoS$_2$ vibrational and light emission properties as a function of layers number deserves a great interest. In particular, it is crucial to evaluate the changes of these properties in the two main steps of Au-assisted exfoliation, i.e. on as-exfoliated MoS$_2$ on Au and after transfer to the final insulating substrate.

In this paper, the evolution of Raman and photoluminescence (PL) spectra of large-area MoS$_2$ (firstly exfoliated on Au and subsequently transferred on an insulating Al$_2$O$_3$/Si substrate) was investigated as a function of the number of layers, evaluated by atomic force microscopy (AFM). We found that the separation $\Delta\omega$ between the in-plane (E$_{2g}$) and out-of-plane (A$_{1g}$) vibrational modes, typically used to estimate MoS$_2$ thickness, exhibits an anomalous large value ($\Delta\omega\sim21.2$ cm$^{-1}$) for 1L MoS$_2$ on Au as compared to 1L MoS$_2$ transferred on Al$_2$O$_3$ ($\Delta\omega\sim18.5$ cm$^{-1}$). Such substrate-related difference was found to gradually decrease while increasing the number of MoS$_2$ layers. Furthermore, PL spectra for 1L MoS$_2$ on Au exhibit a strong quenching and an overall red-shift of the main emission peak at 1.79 eV, compared to the 1.84 eV peak for 1L MoS$_2$ on Al$_2$O$_3$. Such red shift was explained in terms of a higher trion/exciton intensity ratio, probably due to the higher polarizability of the metal substrate, as well as to the smaller equilibrium distance at MoS$_2$/Au interface.

The Au samples employed for the exfoliation were prepared by sequential deposition of 10 nm Ni adhesion layer and 15 nm Au film on a SiO$_2$(900 nm)/Si substrate with DC magnetron sputtering (Quorum Q300TDPLUS). A bulk $MoS_2$ stamp obtained by a freshly cleaved 2H-$MoS_2$ crystal was pressed on the Au substrate immediately after sputtering to prevent the adsorption of contaminants on the Au surface, that could reduce the exfoliation yield [6]. The 2H-$MoS_2$ membrane exfoliated on Au was finally transferred onto an insulating substrate, consisting of 100 nm $Al_2O_3$ deposited on Si. The transfer procedure consisted in the transfer of the Au/$MoS_2$ stack on the $Al_2O_3$ surface and the final etching of Au with a $KI/I_2$ solution [19].

The thickness of $MoS_2$ was evaluated by tapping mode Atomic Force Microscopy (AFM) using a DI3100 equipment by Bruker. The morphology and phase images were acquired simultaneously using sharp silicon tips with a curvature radius of 5 nm. Micro-Raman (μ-R) and micro-Photoluminescence (μ-PL) spectra were obtained using a Horiba Raman system with a confocal microscope (100×) and a laser excitation wavelength of 532 nm. The laser power was filtered with a neutral density (ND) filter at 1% for both spectroscopy methods. A grating of 1800 lines/mm was used to acquire Raman spectra meanwhile a grating of 600 lines/mm to acquire PL spectra. All the spectra were calibrated with respect to the Si peak at 520.7 $cm^{-1}$.

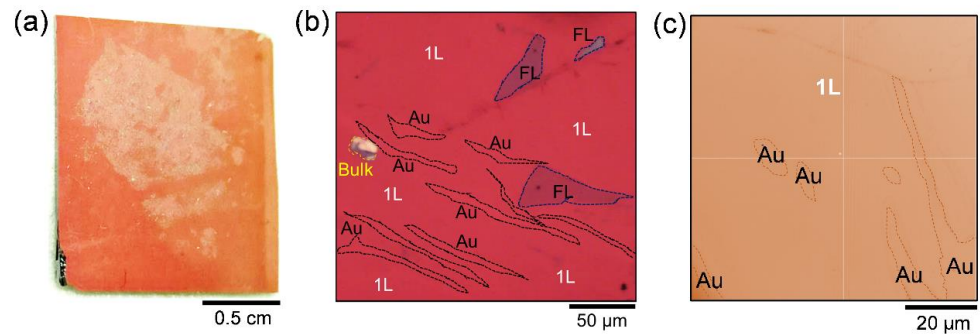

**Figure 1**. (a) Photograph and (b-c) optical microscopy images at two different magnifications of the large area $MoS_2$ membrane on the Au substrate.

Figure 1(a) shows a photograph of the cm$^2$-scale MoS$_2$ exfoliated on the Au substrate, whereas two optical microscopy images at different magnifications are reported in Fig.1(b) and (c), respectively. The variable optical contrast reveals that the MoS$_2$ membrane is predominantly composed by 1L areas, with the presence of FL regions (violet color) and some Au uncovered areas.

After a preliminary identification of 1L and FL areas in the exfoliated MoS$_2$ on Au by observation of the optical contrast, the number of layers was precisely evaluated by tapping mode AFM.

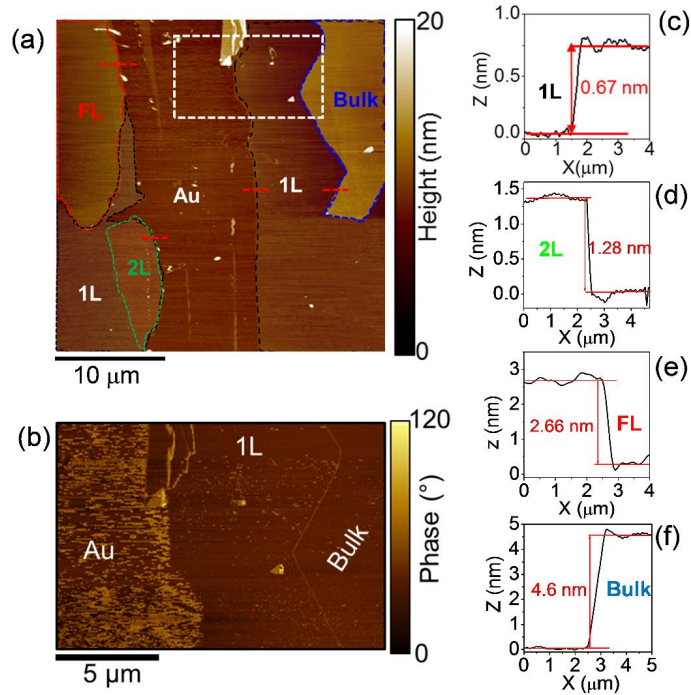

**Figure 2**. (a) Morphological AFM image of a region of the exfoliated MoS$_2$ membrane on Au, containing areas with different MoS$_2$ thicknesses (1L, 2L, FL and bulk) and bare Au areas. (b) Phase image corresponding to the white dashed rectangular region in (a), showing a different contrast between bare Au and MoS$_2$ covered regions. Height line scans from (c) 1L, (d) 2L, (e) FL and (f) bulk MoS$_2$ areas in panel (a).

Fig.2(a) shows an AFM image collected in a region comprising both a bare Au area and $MoS_2$ covered regions with 1L, 2L, FL and multi-layer (bulk) thickness, indicated by different color dashed lines. A phase map collected in the rectangular region indicated by the white dashed line is also reported in Fig.2(b). This image is complementary to the morphology, as it provides a clear identification of the bare Au areas with respect to the $MoS_2$ covered ones, thanks to the very different phase contrast. Fig.2(c) shows a representative height line-scan across the region partially covered by 1L $MoS_2$, from which a ~0.67 nm step was evaluated, consistent with the nominal monolayer thickness of 0.65 nm [1]. Furthermore, the 2L, FL and bulk thicknesses of the different areas in the morphological image are confirmed by the line-scans reported in Fig.2(d), (e) and (f), respectively. Preliminary optical contrast inspection followed by AFM analyses was also employed to identify regions with different thickness in the $MoS_2$ membranes transferred onto the $Al_2O_3$/Si substrate.

In the following, the impact of the two different substrates (Au and $Al_2O_3$) on the vibrational and optical emission properties of $MoS_2$ areas with different thickness has been investigated by μ-R and μ-PL spectroscopy. Fig.3(a) and (b) report a comparison of typical Raman spectra collected on 1L, 2L, FL and bulk regions of the $MoS_2$ membranes exfoliated on Au (a) and transferred onto $Al_2O_3$ (b). Here, the FL region corresponds to 4 layers of $MoS_2$, while the bulk region was formed by 10 layers of $MoS_2$. The characteristic in-plane ($E_{2g}$) and of the out-of-plane ($A_{1g}$) vibrational modes are observed in the spectral range from 370 to 420 $cm^{-1}$. All the spectra were normalized with respect to the $A_{1g}$ peak intensity. Furthermore, vertical dashed lines, corresponding to the $E_{2g}$ and $A_{1g}$ peak positions for 1L $MoS_2$ on Au and $Al_2O_3$, have been reported as a guide for the eye in Fig.3(a) and (b). It can be observed how both the individual peak positions and their separation exhibit a very peculiar dependence on the kind of substrate. While a value of $\Delta\omega \approx 18.5$ $cm^{-1}$ is measured for 1L $MoS_2$ transferred onto $Al_2O_3$, in the case of 1L $MoS_2$ exfoliated on Au the $E_{2g}$ and $A_{1g}$ peaks exhibit a significant red and blue shift, respectively, resulting in a larger value of $\Delta\omega \approx 21.2$ $cm^{-1}$. Furthermore, a different behaviour of the in-plane and out of plane vibrational modes is observed on the two different substrates with increasing the number of layers, as shown in Fig.3(c).

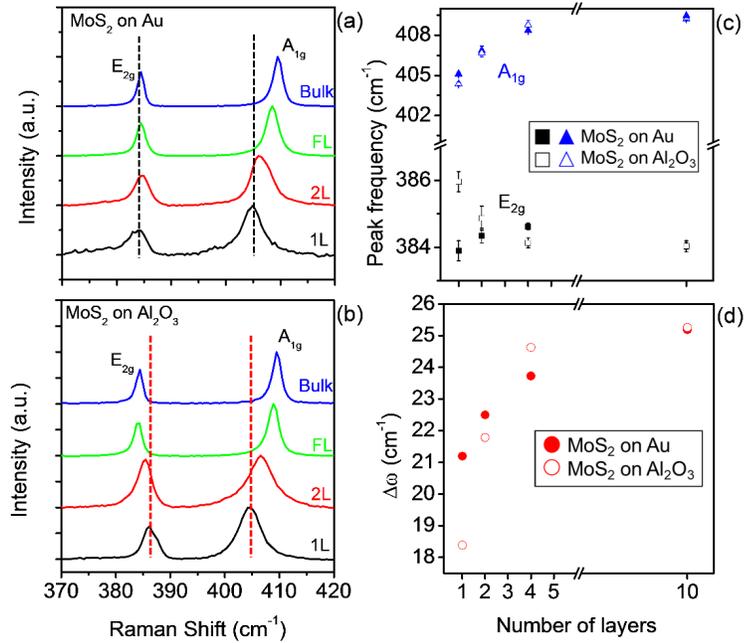

**Figure 3.** Typical Raman spectra of exfoliated $MoS_2$ on the Au substrate (a) and transferred onto $Al_2O_3/Si$ (b) measured on 1L, 2L, FL and bulk $MoS_2$ regions. The black (red) dashed lines indicate the $E_{2g}$ and $A_{1g}$ peaks frequencies for 1L $MoS_2$ on Au ($Al_2O_3$). (c) Behavior of the $E_{2g}$ and $A_{1g}$ peak frequencies as a function of the number of layers for $MoS_2$ on Au (filled squares and triangles) and for $MoS_2$ on $Al_2O_3$ (open squares and triangles). (d) Plot of the peaks frequency difference $\Delta\omega$ as a function of the number of layers for $MoS_2$ on Au (filled red circles) and $MoS_2$ on $Al_2O_3$ (empty red circles).

For both substrates, the $A_{1g}$ peak frequencies (filled and empty triangles) exhibit a similar increasing trend with increasing the $MoS_2$ thickness. In particular, for thin $MoS_2$ membranes (1L-4L) the $A_{1g}$ peak on $MoS_2/Au$ (filled triangles) is slightly blue shifted with respect to $MoS_2/Al_2O_3$ (empty triangles), whereas the two frequencies converge to the same value for bulk samples. On the other hand, the $E_{2g}$ peak frequencies (filled and empty squares) show very different trends on the two substrates. While the decreasing $E_{2g}$ peak frequency with increasing the $MoS_2$ thickness on $Al_2O_3$ (empty squares) is fully coherent with the reported literature results for $MoS_2$ on insulating substrates [20], this peak exhibits an anomalous behavior in the Au case (filled squares). In fact, for 1L $MoS_2$ on Au the $E_{2g}$ is significantly red-shifted (by ~2 $cm^{-1}$) with respect to 1L $MoS_2$ on $Al_2O_3$. Its frequency increases from 1L to 2L $MoS_2$ on Au and remains almost constant for thicker membranes. Noteworthy, for bulk samples, the $E_{2g}$ peak frequencies exhibit the same values on the two substrates. Figure 3(d) shows an increasing behavior of the peaks frequency difference $\Delta\omega$ as a function of the number of $MoS_2$ layers for the two different substrates. Furthermore, starting from a significantly larger value of $\Delta\omega \approx 21.2$ $cm^{-1}$ for 1L $MoS_2$ on Au as compared to $\Delta\omega \approx 18.5$ $cm^{-1}$ for 1L $MoS_2$ on $Al_2O_3$, the difference between the measured $\Delta\omega$ values is gradually reduced with increasing the number of layers, reaching approximately the same value of ~25 $cm^{-1}$ for bulk samples.

It is worth mentioning that the measured $\Delta\omega$ value in the Raman spectra of $MoS_2$ is generally taken as a straightforward way to estimate the number of layers. In particular, for 1L $MoS_2$ exfoliated/grown on common insulating substrates (such as $SiO_2$) the reported values of the separation $\Delta\omega$ between $E_{2g}$ and $A_{1g}$ vibrational peaks can range from ~18 $cm^{-1}$ to ~20 $cm^{-1}$ [20]. Hence, the value of 18.5 $cm^{-1}$ for our 1L $MoS_2$ exfoliated on Au and transferred to the $Al_2O_3$/Si substrate is in the range of the commonly reported literature values. In particular, it is very close to the value measured on 1L $MoS_2$ flakes directly exfoliated on $Al_2O_3$ [21]. On the other hand, for as-exfoliated 1L $MoS_2$ on Au an anomalously large value of $\Delta\omega$=21.2 $cm^{-1}$ is measured. Since the Au-assisted exfoliation is a very clean process (simply achieved by pressing the fresh surface of the bulk $MoS_2$ stamp onto the as-deposited Au film), the large $\Delta\omega$ value cannot be explained by the presence of impurities at $MoS_2$/Au interface or on $MoS_2$ surface. On the other hand, its origin is the strong interaction between $MoS_2$ and the Au substrate [19].

The interaction with the substrate, particularly relevant in the case of ultra-thin $MoS_2$ membranes, can result both in doping effects, associated to charge transfer phenomena, and in tensile or compressive strain effects. The $E_{2g}$ and $A_{1g}$ Raman modes are known to be related to the strain (ε) and doping (n) of $MoS_2$ membranes. In particular, a quantification of the strain type (tensile/compressive) and percentage, as well as of the doping type and carrier density induced on 1L $MoS_2$ by the gold and $Al_2O_3$ substrates have been carried out using the following equations:

$$\omega_{E_{2g}} = \omega_{E_{2g}}^0 - 2\gamma_{E_{2g}} \omega_{E_{2g}}^0 \varepsilon + k_{E_{2g}} n \tag{1}$$

$$\omega_{A_g} = \omega_{A_g}^0 - 2\gamma_{A_g} \omega_{A_g}^0 \varepsilon + k_{A_g} n . \tag{2}$$

Here, $\gamma_{E2g}$=0.68 and $\gamma_{A1g}$=0.21 are the two Grüneisen parameters, correlating the strain ε and the $E_{2g}$ and $A_{1g}$ peaks positions for 1L $MoS_2$ [23], while $k_{E2g}$ =-0.33×10$^{-13}$ cm and $k_{A1g}$ =-2.2×10$^{-13}$ cm are the shift rates of the Raman peaks as a function of the electron density $n$ (in cm$^{-2}$) [22]. Furthermore, $\omega^0_{E2g}$=385 cm$^{-1}$ and $\omega^0_{A1g}$=405 cm$^{-1}$ are the literature values of the $E_{2g}$ and $A_{1g}$ peaks frequencies for a suspended 1L $MoS_2$ membrane under 532 nm excitation [23] , which represents the best approximation of an ideally unstrained and undoped 1L $MoS_2$. According to Eqs. (1) and (2), a biaxial tensile strain ε≈0.21% and a p-type doping $n$≈-0.25×10$^{13}$ cm$^{-2}$ were estimated for 1L $MoS_2$ exfoliated on Au, which was converted into a biaxial compressive strain ε≈-0.25% and n-type doping $n$≈0.5×10$^{13}$ cm$^{-2}$ after transfer to the $Al_2O_3$ substrate. Such n-type behaviour is consistent with the unintentional doping type commonly reported for exfoliated or CVD-grown $MoS_2$, which has been associated to the presence of defects (e.g. sulphur vacancies) or other impurities in the $MoS_2$ lattice [24]. In the case of 1L $MoS_2$ on Au, a strong electron transfer to the substrate is guessed, which overcompensates the native n-type doping, resulting in a net p-type behaviour. Furthermore, the tensile strain for 1L $MoS_2$ on Au can be ascribed to the lattice mismatch between $MoS_2$ and the Au surface, mostly exposing (111) orientation [25,26].

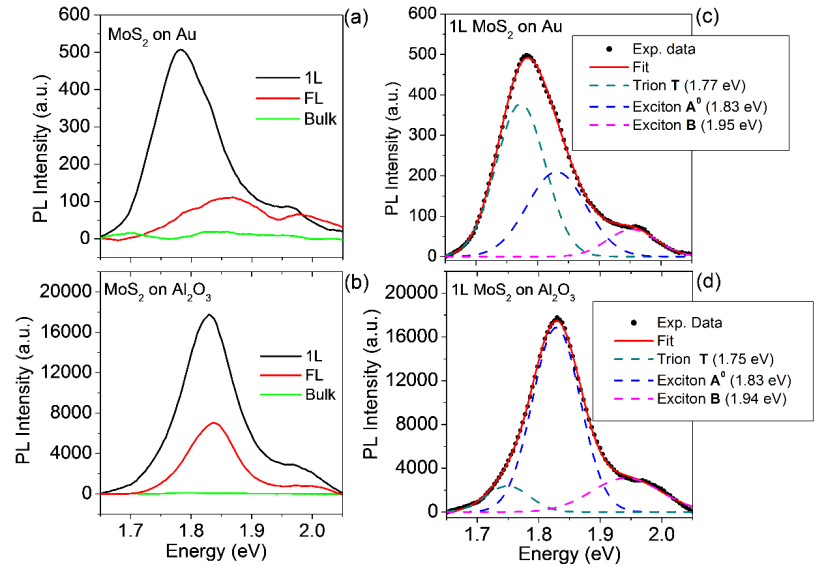

**Figure 4.** Photoluminescence spectra under 532 nm excitation for 1L, FL and bulk $MoS_2$ on Au (a) and $Al_2O_3$ substrate (b). Deconvolution of PL spectra for (c) 1L $MoS_2$ on Au and (d) 1L $MoS_2$ on $Al_2O_3$. Three different components were identified: the trion peak T, the exciton peak $A^0$ and the exciton peak B.

Besides influencing the vibrational properties, the interaction with the substrate is expected to have an impact also on the optical emission behavior of $MoS_2$. Fig.4(a) and (b) illustrate the results of μ-PL analyses performed at room temperature under 532 nm excitation on $MoS_2$ areas with different thickness on the Au and $Al_2O_3$ substrates. In particular, the black, red and green lines represent the PL spectra for 1L, FL and bulk $MoS_2$. In order to perform a reliable comparison of the PL signal on the two different substrates, for each spectrum the intensities were normalized to the intensity of the $MoS_2$ Raman peaks. In this way, the comparison of the PL intensities between Fig.4(a) and (b) demonstrates a quenching (~4 times) of the emission yield for 1L $MoS_2$ exfoliated on Au as compared to 1L $MoS_2$ transferred onto $Al_2O_3$. A reduction of the relative PL intensities when increasing the $MoS_2$ thickness from 1L to FL was consistently observed on both kinds of substrates, with the intensity approaching to zero in bulk samples according to the indirect bandgap. Looking more in details at PL emission for 1L $MoS_2$, for both substrates the PL spectra exhibit a main intense peak at lower energy and another weaker peak at higher energy, associated to the $MoS_2$ band splitting due to spin-orbit coupling [27,28]. While the main peak for 1L $MoS_2$ on $Al_2O_3$ is located at 1.84 eV, similarly to what typically reported on other insulating substrates [29], a significant red shift to 1.79 eV is observed for 1L $MoS_2$ on the Au substrate. To get a deeper insight in the PL emission mechanisms of 1L $MoS_2$ on the two different substrates, a deconvolution of the two representative spectra has been carried out, as reported in Fig.4(c) and (d). In both cases, the best fit was obtained considering three Gaussian peaks, which were associated to a trionic contribution T (green dashed line), and two excitonic contributions, i.e. the exciton $A^0$ (blue) and the exciton B (grey) [30,31,32]. Differently from neutral excitons, consisting of a bound electron/hole pair, trions are charged quasiparticles formed by two electrons and a hole [30]. Noteworthy, while the exciton peak $A^0$ at 1.84 eV represents the main PL contribution for 1L $MoS_2$ on $Al_2O_3$, the trion peak T at 1.78 eV appears to be the dominant one in the case of 1L $MoS_2$ on Au. Finally, the B exciton peak at 1.94 eV for 1L $MoS_2$ on $Al_2O_3$ exhibits a significantly higher full width at half maximum (FWHM) with respect to the corresponding peak (at 1.96 eV) for 1L $MoS_2$ on Au. As indicated in the labels of Fig.4(c) and (d), the T, $A^0$ and B peaks obtained by the deconvolution are slightly blue-shifted in the case of 1L $MoS_2$ on Au with respect to 1L $MoS_2$ on $Al_2O_3$. However, the overall red shift of the PL spectra for 1L $MoS_2$ on Au is due to the higher intensity of the trion contribution. As reported in recent theoretical studies [33,34], this effect can be ascribed to the high polarizability of the metal substrate and to the low $MoS_2$/Au equilibrium distance enhancing the trion population and at the same time quenching the overall emission amplitude [6]. Noteworthy, in the case of FL $MoS_2$ on Au (Fig.4(a)) the main PL peak appears to be broader and blue-shifted with respect to the monolayer one, and its energy is closer to that of FL $MoS_2$ on $Al_2O_3$. This observation suggests that the increase of the $MoS_2$ thickness results in a reduced effect of the substrate not only on vibrational properties but also on PL emission.

In conclusion, we have deeply investigated the substrate effects on the Raman and PL emission properties of $cm^2$ – wide $MoS_2$ membranes exfoliated on Au and subsequently transferred on an insulating $Al_2O_3$/Si substrate. For as-exfoliated 1L $MoS_2$ on Au, Raman spectra showed an anomalous large value of $\Delta\omega\approx21.2$ $cm^{-1}$ (due to the tensile strain and p-type doping induced by the substrate) as compared to the typical one (~18.5 $cm^{-1}$) measured after the transfer of 1L $MoS_2$ on $Al_2O_3$ and complete removal of Au. Such substrate-related differences, were found to gradually decrease while increasing the number of $MoS_2$ layers. These results have also practical implications, indicating that Raman spectroscopy should be used in combination with other physical characterizations (e.g. AFM or transmission electron microscopy) to unambiguously evaluate the number of $MoS_2$ layers. Furthermore, PL spectra for 1L $MoS_2$ on Au exhibit a strong quenching and an overall red-shift of the main emission peak at 1.79 eV, compared to the 1.84 eV peak position for 1L $MoS_2$ on $Al_2O_3$. Such red shift was explained in terms of a higher trion/exciton intensity ratio, indicating how the relative population of quasiparticles generated under light excitation is significantly affected by the 1L $MoS_2$/Au interaction.

These results will be relevant in view of the widespread applications of large-area $MoS_2$ membranes produced by the gold-assisted exfoliation in electronics and optoelectronics.

S. Di Franco (CNR-IMM) is acknowledged for the expert technical assistance in samples preparation. This work has been supported, in part, by MUR in the framework of the FlagERA-JTC2019 project ETMOS. Part of this work has been carried out in the Italian Infrastructure *Beyond-Nano Upgrade*.

The data that support the findings of this study are available from the corresponding author upon reasonable request.

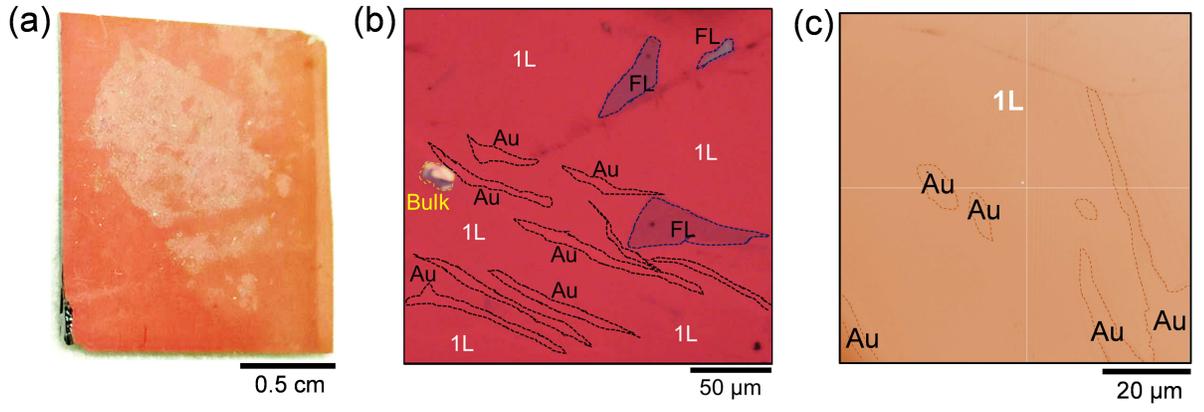

Applied Physics Letters

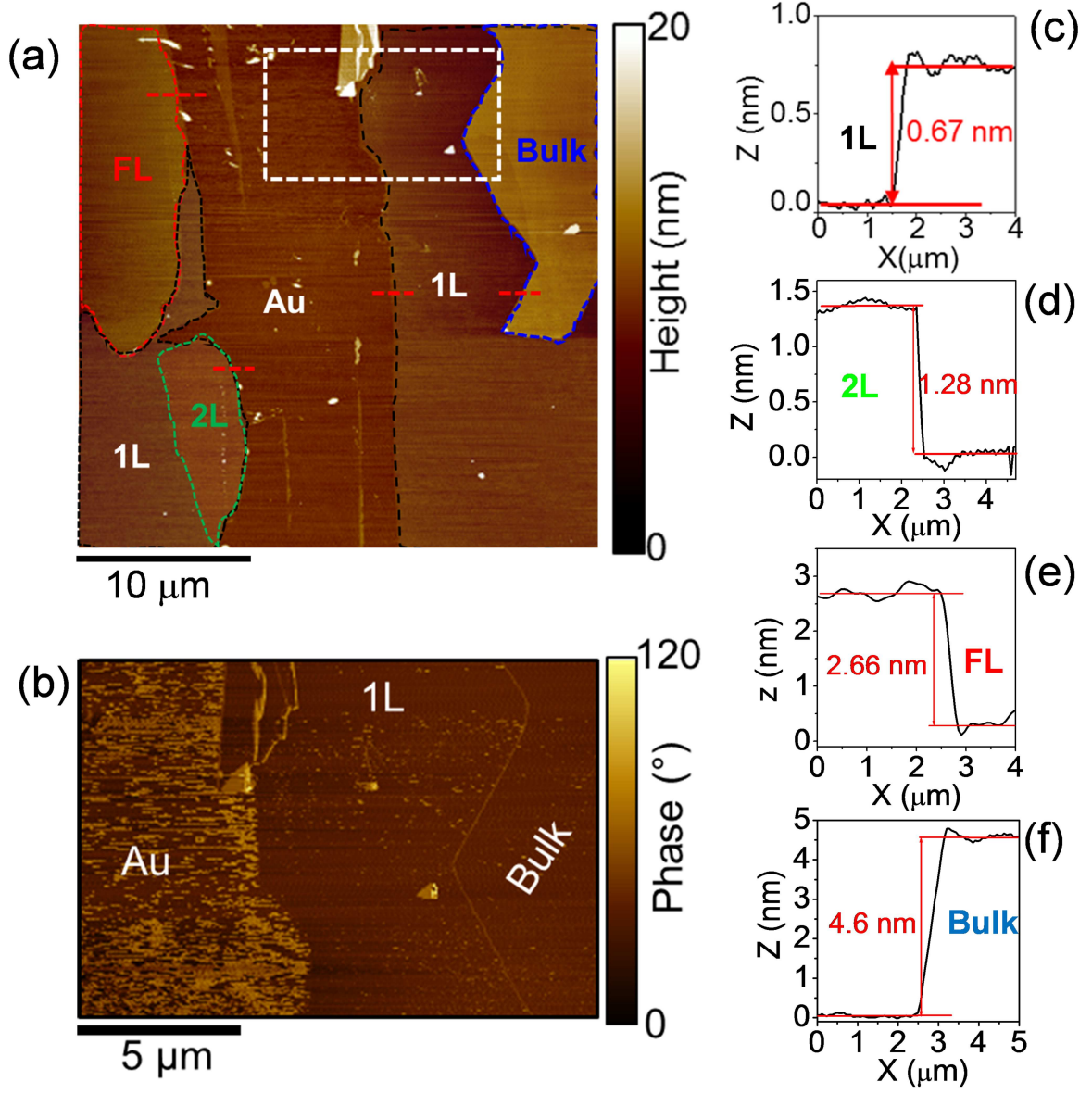

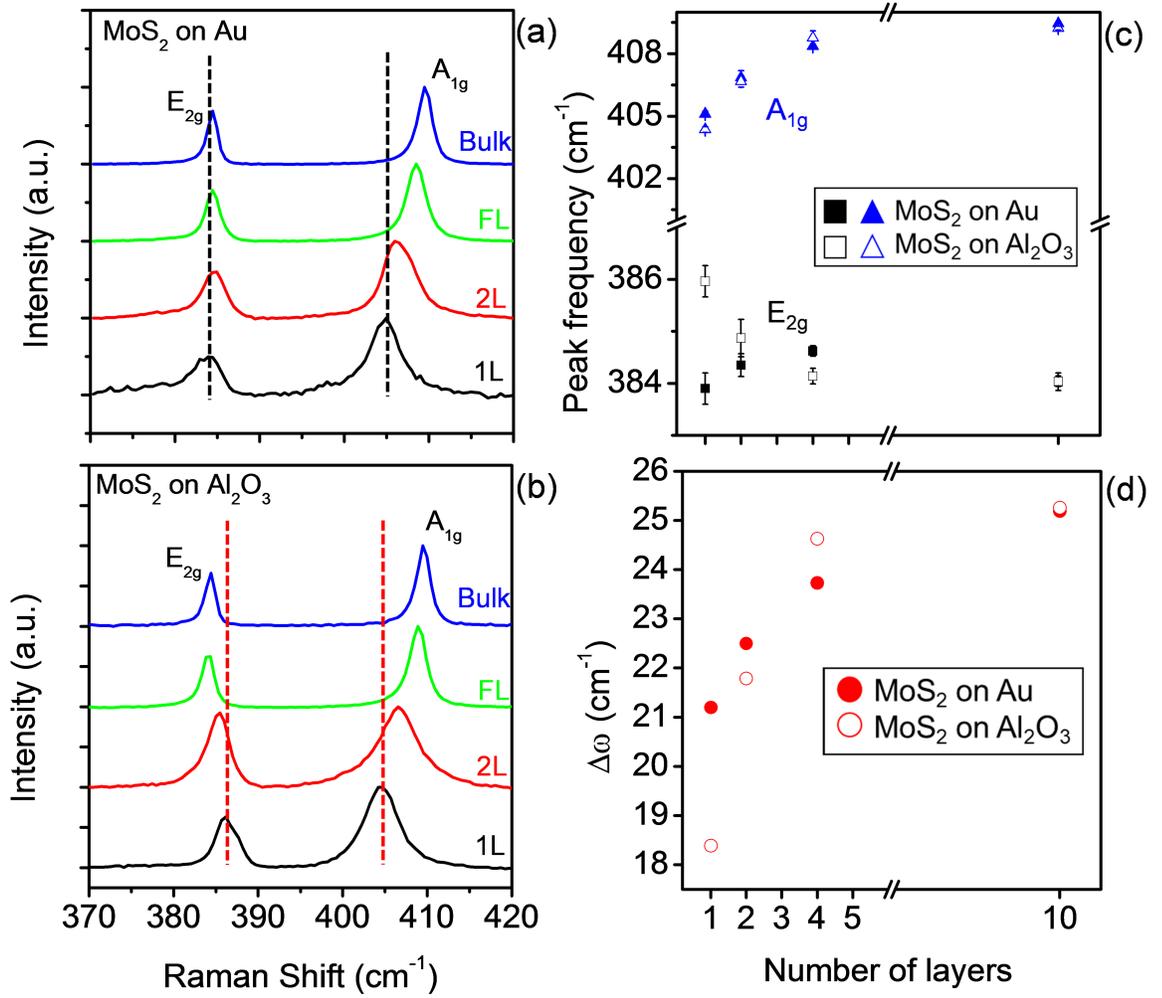

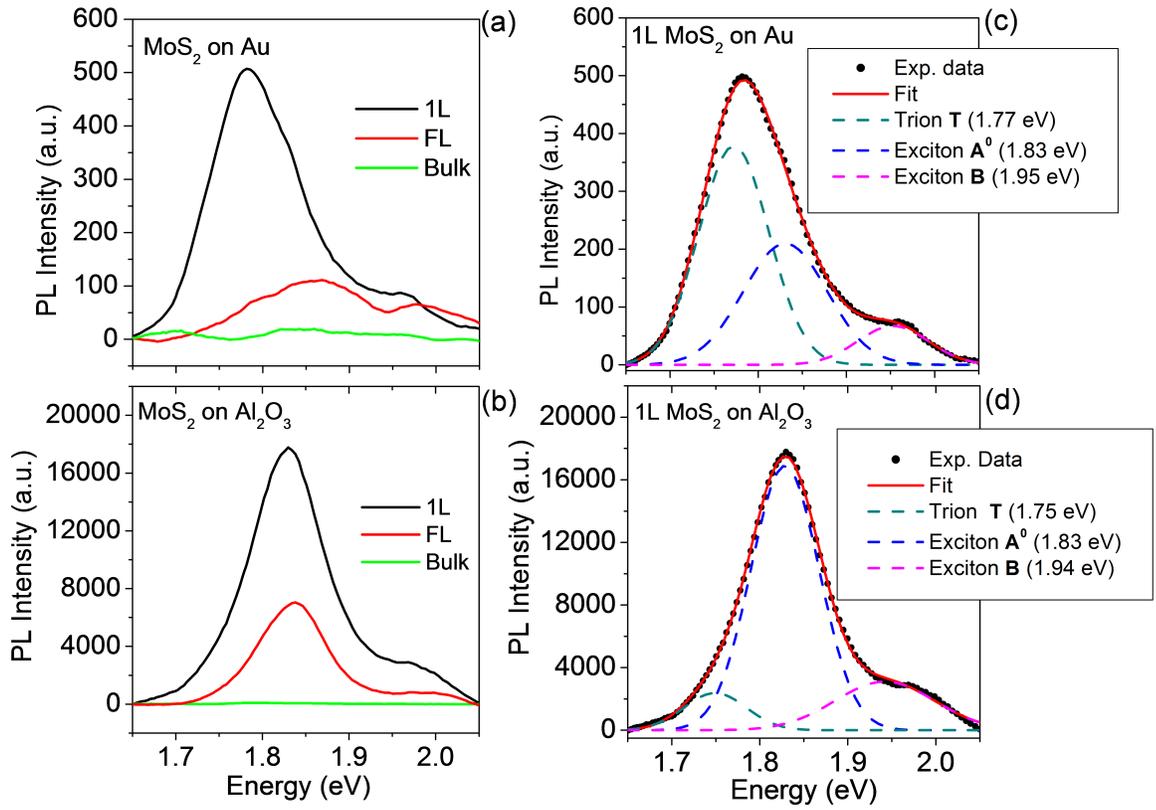